\documentclass[doublecol]{epl2} 
\usepackage{graphicx}
\usepackage{amsmath,amssymb,revsymb}
\usepackage{color}

\def\bfB{\mbox{\boldmath $B$}}
\def\bfu{\mbox{\boldmath $u$}}
\def\bfnabla{\mbox{\boldmath $\nabla$}}
\def\bfpsi{\mbox{\boldmath $G$}}

\title{A numerical model of the VKS experiment } \author{Christophe
  J.P. Gissinger\inst{1,2}} \shortauthor{Gissinger} \institute{
  \inst{1} Laboratoire de Physique Statistique, \'Ecole Normale
  Sup\'erieure CNRS, 24 rue Lhomond, F-75005 Paris
  (France).\\ \inst{2} Laboratoire de Radiostronomie, \'Ecole Normale
  Sup\'erieure CNRS, 24 rue Lhomond, F-75005 Paris (France).\\}

\pacs{91.25.Cw}{Origins and models of the magnetic field; dynamo theories}
\pacs{47.65.+a}{Magnetohydrodynamics and electrohydrodynamics}

\abstract{We present numerical simulations of the magnetic field
  generated by the flow of liquid sodium driven by two
  counter-rotating impellers (VKS experiment). Using a kinematic code
  in cylindrical geometry, it is shown that different magnetic modes
  can be generated depending on the flow configuration. While the time
  averaged axisymmetric mean flow generates an equatorial dipole, our
  simulations show that an axial field of either dipolar or
  quadrupolar symmetry can be generated by taking into account
  non-axisymmetric components of the flow. Moreover, we show that by
  breaking a symmetry of the flow, the magnetic field
  becomes oscillatory. This leads to reversals of the axial dipole
  polarity, involving a competition with the quadrupolar component.}

\begin{document}

\maketitle

\section{Introduction}
The dynamo effect is a process by which a magnetic field is generated
by the flow of an electrically conducting fluid. It is believed to be
responsible for magnetic fields of planets, stars and galaxies
\cite{moffatt}. Fluid dynamos have been observed in laboratory
experiments in Karlsruhe \cite{karlsruhe} and Riga \cite{riga}. More
recently, the VKS experiment displayed self-generation in a less
constrained geometry, i.e., a von K\'arm\'an swirling flow generated
between two counter-rotating disks in a cylinder \cite{monchaux07}. In
contrast with Karlsruhe and Riga experiments, the observed magnetic
field strongly differs from the one computed taking into account the
mean flow alone. Previous simulations, using the mean flow (time
averaged) of the VKS experiment or an analytical velocity field with
the same geometry, predicted an equatorial dipole
\cite{marie03,bourgoin04,gissinger08a,stefani06} in contradiction with
the axial dipole observed in the experiment. Understanding the
geometry of the magnetic field observed in the experiment is still an
open problem. In addition, time-dependent regimes, including field
reversals are observed when the impellers rotate at different
frequencies \cite{berhanu07}.  No numerical study of this effect have
been performed so far. We address these problems using a kinematic
dynamo code in a cylindrical geometry. By considering time dependent
and non-axisymmetric fluctuations of the velocity field, we show that
the system is able to generate a nearly axisymmetric dipolar
field. Another result of this study is that when the analytic flow
mimics two disks counter-rotating with different frequencies, the
system bifurcates to a regime of oscillations between dipole and
quadrupole, illustrating a recent model proposed in \cite{Petrelis08}
in order to explain reversals of the magnetic field in the VKS
experiment or in the Earth.  We will see that an $\alpha-\omega$
mechanism can explain the generation of the axial field and we
understand the transition to oscillations as a saddle node bifurcation
associated with the breaking of a symmetry in the flow.

\section{Numerical model} 
In the VKS experiment, a turbulent von K\'arm\'an flow of liquid
sodium is generated by two counter-rotating impellers (with rotation
frequencies $F_1$ and $F_2$). The impellers are made of iron disks of
radius $154$ mm, fitted with $8$ iron blades of height $41.2$ mm, and
are placed $371$ mm apart in an inner cylinder of radius $206$ mm and
length $524$ mm. It is surrounded by sodium at rest in another
concentric cylindrical vessel, $578$ mm in inner diameter.  The
magnetic Reynolds numbers are defined as $R_{mi} = 2 \pi \mu_0 \sigma
R^2 F_i$ where $\mu_0$ is the magnetic permeability of vacuum.  When
the impellers are operated at equal and opposite rotation rates $F$, a
statistically stationary magnetic field with either polarity is
generated above $R_m \sim 30$~\cite{monchaux07}. The mean field
involves an azimuthal component and a poloidal one which is dominated
by an axial dipole.

When the disks are counter-rotating at the same frequency, the
structure of the mean flow (averaged in time) has the following
characteristics: the fluid is ejected radially from the disks by
centrifugal force and loops back towards the axis in the mid-plane
between the impellers. A strong differential rotation is superimposed
on this poloidal flow, which generates a high shear in the
mid-plane. Because of the axisymmetry of this flow, we expect from
Cowling's theorem that an axisymmetric magnetic field can not be
generated. Simulations based on the mean flow indeed generate a
non-axisymmetric equatorial dipole
\cite{marie03,bourgoin04,gissinger08a}. A better description of the
VKS experiment clearly needs to involve the non-axisymmetric
components of the flow. In this perspective, the geometry of the
experimentally observed field has been understood with a simple model
of an $\alpha-\omega$ dynamo \cite{petrelis07}. In the VKS experiment,
the strong differential rotation is very efficient to convert poloidal
into toroidal magnetic field, via an $\omega$ effect. In addition, the
flow that is ejected by the centrifugal force close to each impeller
is strongly helical due to the vortices created between the successive
$8$ blades. This non-axisymmetric helical flow drives an $\alpha$
effect which converts the toroidal field in a poloidal one. These two
effects have been proposed as being responsible for the magnetic field
generation, which thus results from an $\alpha-\omega$ dynamo. In the
present study, we consider an analytic test velocity $\bfu$ taking
into account the time dependent $m=8$ structure due to the blades as
follows:

\begin{equation}
\bfu = {\bf U} + \bfnabla \times \left(\kappa{\bfpsi}\right),
\label{flow}
\end{equation} 
where ${\bf U}$ mimics the mean flow. It is given in cylindrical
coordinates $(r , \phi, z)$ by
 \begin{eqnarray}
 U_{r}&=&-{\pi\over 2}r{\left(1-r\right)}^2\left(1+2r\right)\cos\left(\pi z\right) ,\\
 U_{\phi}&=&{8\over \pi}r\left(1-r\right) \arcsin\left(z\right) ,\\
 U_{z}&=&\left(1-r\right)\left(1+r-5r^2\right) \sin\left(\pi z\right). 
 \label{flow2}
 \end{eqnarray}

The vector potential $\bfpsi$ describes the non-axisymmetric
fluctuations due to the blades of the disks. A simple way to represent
these vortices is to take :
\begin{multline}
G_{r}=\left[1+\cos\left(m\phi-\omega_1 t\right)\right]r\sin(\pi r)e^{-\zeta(z-1)^2}\left(z-1\right)\\ +\left[1+\cos\left(m\phi+\omega_2 t\right)\right]r\sin(\pi r)e^{-\zeta(z+1)^2}\left(z+1\right),                    
\end{multline}
\begin{equation}
G_{\phi}=-{3}r^2{\left(1-r\right)}^2\left(1+2r\right)z^2\sin\left(\pi z\right)\cos(m\phi\pm\omega_i t).
\end{equation}

In this expression, $G_{r}$ is related to a flow in the ($z,\phi$)
plane and models the $8$ vortices created by the blades of each
disk. These vortices are rotating with the two disks at angular
velocity $\omega_1$ and $\omega_2$. The $G_{\phi}$ function represents
the $\phi$-modulation of the poloidal mean recirculation associated
with these vortices. The $z$ dependence of $G_{r}$ is parametrized by
$\zeta$ and determines the extension of the perturbation close to the
disks. The relative intensity of the mean flow and the
non-axisymmetric perturbation is fixed by the value of $\kappa$. Note
that the system presents an important symmetry: the flow is invariant
by a rotation of an angle $\pi$ around any axis in the equatorial
plane. In the following, we will denote this symmetry by $R_\pi$. In
some simulations, we will describe the situation where one of the
disks is spinning faster than the other one, thus breaking the $R_\pi$
symmetry. As in the symmetric case, there are several ways to
implement this situation in our analytical velocity, and we take the
simplest one. Spinning one disks faster than the other one is
simulated by adding a global rotation ${\bf W}$ with recirculation
which breaks the $R_\pi$ symmetry. In this case, ${\bf U}$ become
${\bf U}+C{\bf W}$, with ${\bf W}$ defined by
\begin{eqnarray}
 W_{r}&=&{\pi\over 4}r{\left(1-r\right)}^2\left(1+2r\right)\sin\left(\pi z/2\right) , \\
 W_{\phi}&=&6r\left(1-r\right) , \\
 W_{z}&=&\left(1-r\right)\left(1+r-5r^2\right) \cos\left(\pi z/2\right) .
\end{eqnarray}
This is the mean flow corresponding to the disk at $z=+1$ spinning
alone. The parameter $C$ thus controls the deviation from exact
counter-rotation ($C=0$), and traces back to the difference between the
two disk frequencies in the VKS experiment. In addition, it is
reasonnable to assume that the non-axisymetric velocity component is also
modified when $C\ne 0$. This is achieved by the simple transformation
$\omega_1=\omega_1(1+C)$ and $\kappa=\kappa (1+C)$ for $z>0$.\\

 Note that the expression for the velocity used here is arbitrary, and
 there are probably several ways to describe with more accuracy the
 von Karman flow. However, the purpose of this study is to show that
 the structure and the dynamics of the magnetic field in the VKS
 experiment can be easily reproduced by taking into account the main
 geometrical properties of the flow, i.e.  the vortical structure near
 the disks and the breaking of the $R_\pi$ symmetry when the disks
 counter-rotate at different frequencies.

We perform direct numerical simulations of the kinematic dynamo
problem, solving the induction equation governing the evolution of the
solenoidal magnetic field $\bf B$
\begin{equation}
\frac{\partial \bfB}{\partial t} = \bfnabla \times \left(
\bfu \times \bfB \right) +\frac{1}{Rm} \Delta \bfB \, ,
\end{equation}
written in dimensionless form using the advective timescale.  The
magnetic Reynolds number $Rm$ is defined as $Rm=\mu _0 \sigma R
U_{max}$, where $R$ is the radius of the impellers and $U_{max}$ is
the peak velocity of the mean flow. The radius of the cylinder is
$L=4R/3$ and the total height is $H=2L$. \\

The above system with the flow given in (\ref{flow}) is
solved using a finite volume code adapted from \cite{teyssier}.  A
filtering of high-frequency modes in the $\phi$ direction has been
implemented to circumvent the severe restriction on the CFL number
induced by cylindrical coordinates near the axis.  Also a centered
second order scheme has been prefered to an up-wind scheme as
resistive effects are here important enough to regularise the
solution.  As in \cite{teyssier}, we ensure that $\bfnabla \cdot
\bfB=0$ is exactly satisfied using a constraint transport algorithm.
The finite volume solver is fully three-dimensional. 
Two types of magnetic boundary conditions are used: insulating
boundaries, using the boundary element formalism as introduced in
\cite{iskakov}, or ferromagnetic boundaries, by forcing the magnetic
field lines to be normal to the external wall, as used in
\cite{gissinger08a}. In all cases we do not include sodium at rest
around the vessel or behind the disks.

\section{Structure of the magnetic field}

\begin{figure}[!htb]
\centerline{\includegraphics[width=9cm]{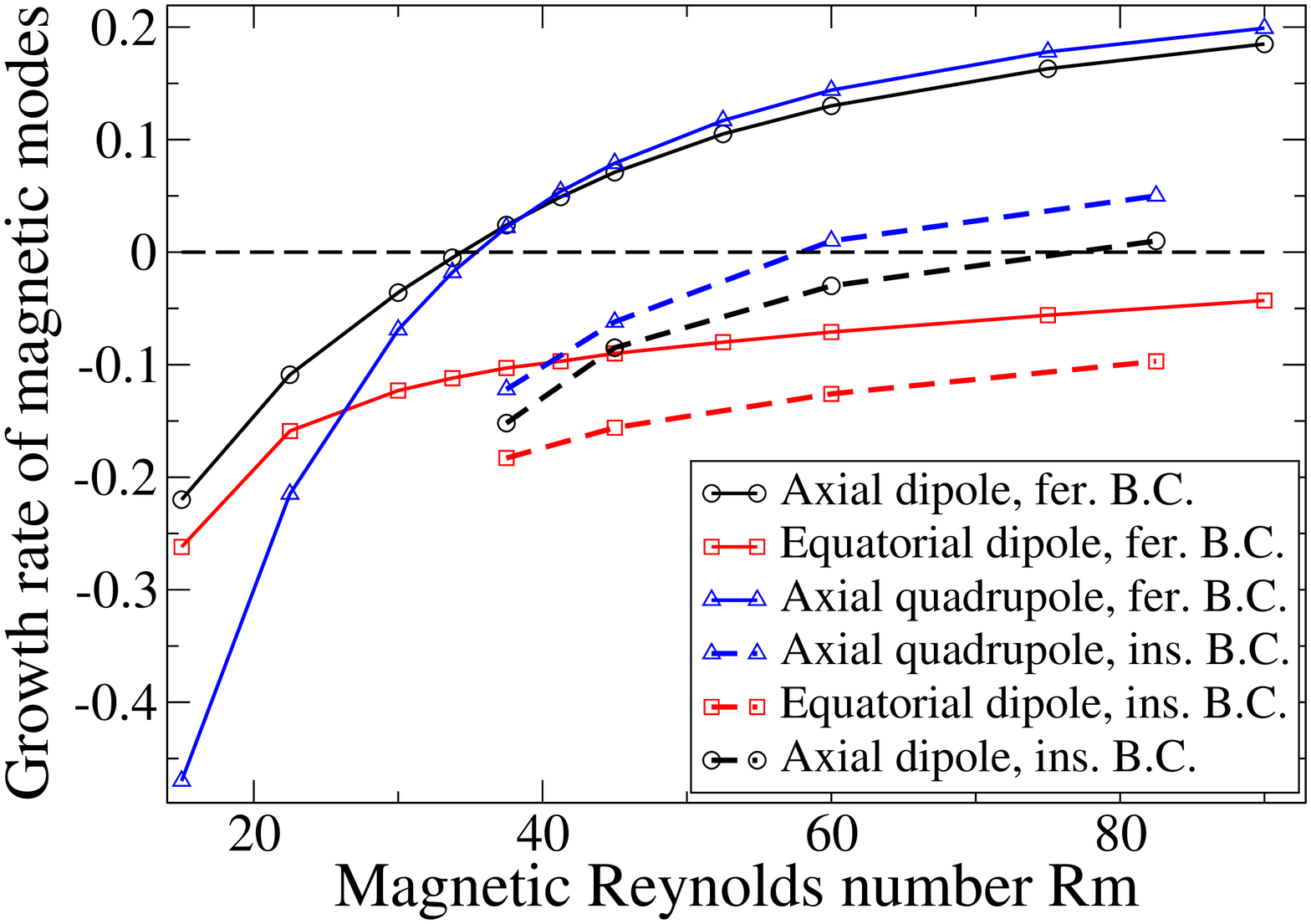}}
\caption{Evolution of the growth rates of the magnetic modes when $Rm$
  is increased in exact counter-rotation. For ferromagnetic
  boundaries, we show the $m=0$ axial dipolar mode
  (circles). Triangles: $m=0$ quadrupolar field. Squares: equatorial
  dipole. Note the close values of the onset for dipole and
  quadrupole. Growth rates with insulating boundaries are also
  displayed (dashed lines), showing much larger onset values for all
  modes. }
\label{rate}
\end{figure}
\begin{figure}[!htb]
\centerline{\includegraphics[width=9cm]{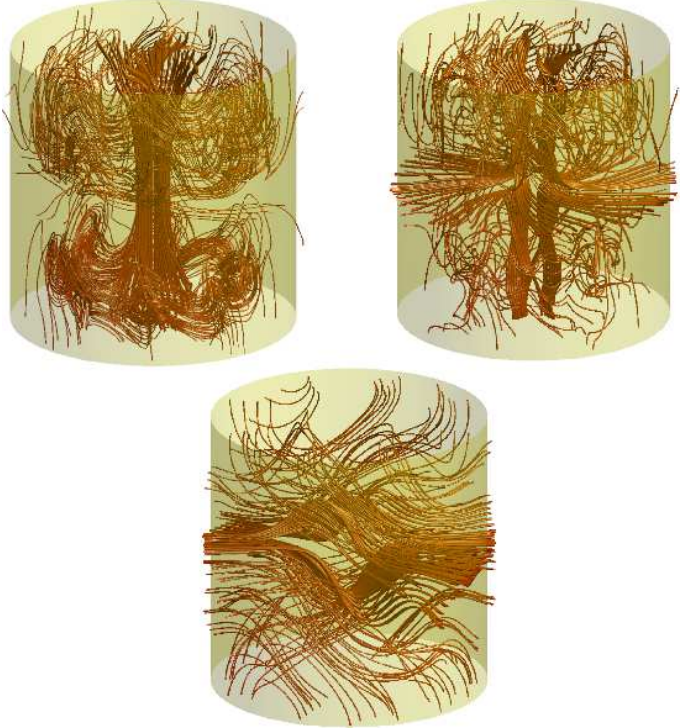}}
\caption{Snapshot of the magnetic field lines of different modes
  obtained in the numerical simulations. Axial dipole (top, left),
  quadrupolar (top, right) and equatorial dipole (bottom). Observe
  that the axial modes always involve $m=8$ component.}
\label{mode}
\end{figure}

The convergence of the numerical implementation has been carefully
validated comparing simulations at different resolutions.  We report
results obtained with a resolution of $150 \times 150 \times 128$ . In
the simulations presented here, we use $\zeta=30$ and $\kappa=1$. This
corresponds to vortices of a typical size of $1/5$ of the total height
of the cylinder, and with a velocity of the same order than the mean
flow. These parameters are comparable to what is expected in the real
experiment. Let us study first the situation corresponding to the
counter-rotating case. Figure ~\ref{rate} shows the growth rate
of different magnetic modes as a function of the magnetic Reynolds
number. We see that the non-axisymetric flow leads to the generation
of an axisymmetric $m=0$ magnetic mode with a dipolar symmetry, which
bifurcates for $Rm\approx 34$, using ferromagnetic boundary
conditions. Because of the $m=8$ structure in the velocity field, this
mode is in fact associated with a $m=8$ magnetic component, growing at
the same rate. However, this small scale structure is time dependent
and averages on a few advective times. In figure
~\ref{rate}, we see that the first unstable mode is an axial dipole
similar to the mean field observed in the VKS experiment, and that the
quadrupolar mode is also unstable for larger $Rm$. The other modes are
not unstable in the range of $Rm$ studied here. In particular, we see
that the non axisymmetric component of the flow strongly inhibits the
$m=1$ equatorial dipole, which bifurcates around $Rm=45$ for
$\kappa=0$, i.e. with the mean flow alone (see figure \ref{mode},
bottom). In general, the two first axial modes always
display similar threshold values, and depending on the parameters of
the flow, the first unstable axisymmetric structure can be either a
dipole or a quadrupole. In the VKS experiment, a quadrupolar field has
never been observed for counter-rotation of the disks at the same
frequency. The different magnetic structures are represented in figure
~\ref{mode}. 

\begin{figure}[!htb]
\centerline{\includegraphics[width=7cm]{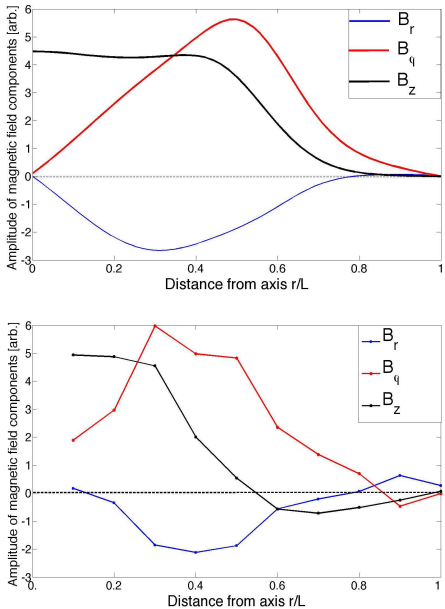}}
\caption{Comparison of the numerical simulations (top) and
  experimental results (bottom). We show the radial profiles for the 3
  components of the axial dipolar field measured near one of the
  disks, when the disks counter-rotate for $Rm=36$. Fields have been
  $\phi$-averaged (simulations) or time averaged (experimental data,
  from \cite{Monchaux09}).}
\label{comp}
\end{figure}

 In figure \ref{comp}, we represent the radial profiles of the
 axisymmetric components of the dipole for $Rm=36$, which are similar
 to the ones of the mean field observed in the VKS experiment
 \cite{Monchaux09}. These results confirm the mechanism proposed in
 \cite{petrelis07}, which states that the non-axisymmetric vortices
 near the disks could be responsible for the generation of the
 observed magnetic field.  In a recent $\alpha$-parametrized mean
 field approach, a completely different result was found, for which
 the $\alpha$ term is said to be several times larger than any
 realistic value based on the VKS experiment \cite{Laguerre08}. Here,
 our study shows a critical $Rm$ comparable to the experiment. In
 addition, the maximum intensity of the vortices compared to the mean
 flow, about a factor $1.5$ here for $\kappa=1$, is
 reasonnable. Figure ~\ref{rate} also shows the effect of the magnetic
 boundary conditions on the dynamo threshold of the different
 modes. We see that using ferromagnetic boundary conditions is very
 efficient for decreasing the dynamo threshold. For insulating
 boundaries, the $m=0$ mode bifurcates for $Rm=58$, whereas its
 threshold is $Rm=34$ for ferromagnetic boundary conditions. Note that
 in the case of insulating boundaries, the first unstable mode is a
 quadrupole, and the axial dipole bifurcates for $Rm=76$.  This
 confirms the role played by soft iron disks in the experiment:
 changing the geometry of the magnetic field lines near the external
 wall yields a strong reduction of the instability threshold. This was
 already observed with the mean flow alone \cite{gissinger08a} and
 could explain why dynamo action has only been observed in the VKS
 experiment when soft iron disks are used.

\section{Dynamics of the magnetic field for different rotation rates}

In the VKS experiment, several dynamical behaviors occur when the
rotation rates of the two disks are differents: periodic oscillations,
bursts and chaotic reversals of the magnetic field have been
reported. When the velocity difference of the disks is increased from
zero, the stationnary dipolar field is first modified by the addition
of a quadrupolar component before diplaying a transition to a time
dependent regime \cite{Ravelet08}.\\

 When the disks counter-rotates at the same frequency, the flow is
 invariant under the $R_\pi$ symmetry. Consequentely, dipolar and
 quadrupolar modes, which transform differently under $R_\pi$, are
 linearly decoupled. We observe in figure \ref{rate} that the dipole
 and quadrupole modes have slightly different growth rates at the
 dynamo onset, the neutral mode being a dipolar mode (see figure
 ~\ref{mode}, bottom). When one disk is spinning faster than the other
 one, the $R_\pi$ symmetry of the flow is broken and dipolar and
 quadrupolar modes get coupled. Consequentely, the growing unstable
 mode has to be a combination of a dipole and a quadrupole. The ratio
 between dipolar and quadrupolar components depends on the intensity of
 the breaking of the $R_\pi$ symmetry.

\begin{figure}[!htb]
\centerline{\includegraphics[width=9cm]{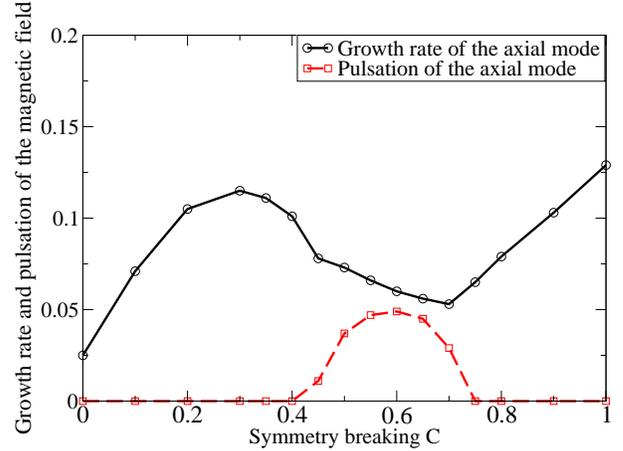}}
\caption{Evolution of the growth rate and pulsation of the axial mode
  when the $R_\pi$ symmetry is broken. For $C<0.4$, the growing mode
  is a combination of a dipole and a quadrupole (growth rate, black
  circles). At $C=0.4$ the system undergoes a bifurcation to
  oscillations (red squares). Note that, in this range, the
  oscillatory regime only occurs in a pocket of the parameter space,
  for $0.4<C<0.75$. }
\label{saddle}
\end{figure}

\begin{figure}[!htb]
\centerline{\includegraphics[width=9cm]{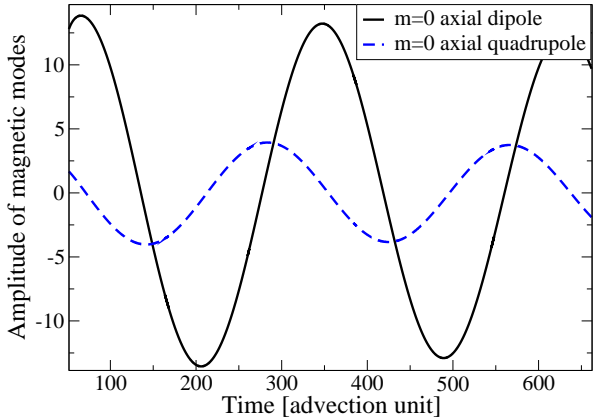}}
\caption{Time evolution of the magnetic field during oscillations.
  The system is very close to criticality, for $Rm=30$ and
  $C=0.6$, such that the growth rate is very small.
  Note the transfer between the axial dipole ( full line) and the quadrupole
   (dashed line).}
\label{time}
\end{figure}

 Figure ~\ref{saddle} shows the evolution of the growth rates of the
 modes when we increase the parameter $C$ representing the departure
 from counter-rotation. We start from $Rm=37$, and $C$ is increased
 from zero to one. Because of the breaking of the $R_\pi$ symmetry
 when $C\ne 0$, the growing mode is not a pure dipole
 anymore. Similarly, the initially quadrupolar mode displays some
 dipolar component. In this linear problem, only the most unstable
 mode is easily obtained from the simulations, since dipolar and
 quadrupolar families are mixed when the $R_\pi$ symmetry is
 broken. We see that the growth rate of the axial mode is changed when
 $C$ is increased. For $C=0.4$, the system bifurcates to an
 oscillatory regime, when two different real eigenvalues transform
 into complex conjugate ones. These results have been recently
 understood in the framework of a simple model, based on a saddle node
 bifurcation \cite{Petrelis08}. If we denote by $D$ the amplitude of
 the eigenmode with dipolar symmetry and by $Q$ the quadrupolar one, a
 simple way to understand these linear results is to write the
 evolution of the modes near the threshold:
\begin{equation}
\dot{D}=\alpha(C) D + \beta(C) Q ,
\end{equation}
\begin{equation}
\dot{Q}=\gamma(C) D + \delta(C) Q.
\end{equation}
where dots stand for time derivation. The eigenvalues $\lambda_i$ of
this system are given by $2\lambda_i=\alpha+\delta \pm
\sqrt{(\alpha-\delta)^2+4\beta\gamma}$. When the disks are spinning at
the same rotation rate, $\beta$ and $\gamma$ vanish and the dipole and
the quadrupole are not linearly coupled, giving two real eingenvalues
$\lambda_1$ and $\lambda_2$. For $C\ne 0$, when $\beta\gamma$ is
negative and sufficiently large, $\lambda_1$ and $\lambda_2$ become
complex conjugate eigenvalues. The system thus bifurcates to an
oscillatory regime, in agreement with the numerical simulations shown
here.  The pulsation of the oscillatory mode is represented in figure
\ref{saddle}. Note that the period of the oscillations diverges at
threshold. An interesting feature is that these periodic reversals of
the magnetic field only occur in a small range of the parameter space,
for $0.4<C<0.75$. This shows that the relation between $C$ and the
parameters of the system are rather complex. This is also the case in
the VKS experiment, where reversals, periodic oscillations and bursts
appear only in small pockets in the parameter space.\\

 Figure \ref{time} displays a typical time evolution of the magnetic
 field during oscillations. Here, the system is investigated very
 close to criticality ($Rm=30$ and $C=1.6$), such that the exponential
 behavior is very weak. These oscillations involve a competition
 between an axial dipole and an axial quadrupole. We see in
 particular that the two components are in quadrature. This means that
 the magnetic field does change shape. The present study being linear,
 it can only illustrate the transition between stationnary and
 oscillatory dynamos. In particular, the present simulations cannot
 reproduce the non-periodic reversals of the magnetic field observed
 in the VKS experiment \cite{berhanu07}. This more realistic situation
 is reported in \cite{gissinger09}, where chaotic reversals of the
 magnetic field are observed in fully turbulent simulations.

\section{Conclusion}

      In this study, we have used a simple model of a von Karman flow to
describe the structure and some of the dynamical behaviors of the
magnetic field in the VKS experiment. In particular, we have shown
that taking into account the non-axisymmetric fluctuations of the flow
leads to the generation of a nearly axisymmetric dipole. This confirms
that the helical flow created by the vortices between the blades of
each disk can be involved in the dynamo process, possibly via an
$\alpha-\omega$ mechanism. This flow generates an axial dipole or
quadrupole, and also inhibits the equatorial dipole. In addition, we
have shown that the presence of ferromagnetic boundaries can strongly
reduce the threshold of the dynamo instability. Another result of this
study concerns the dynamics of the magnetic field. When the disks are
spinning at different rates, the $R_\pi$ symmetry of the flow is
broken. Our simulations show that this can lead to an oscillatory
regime between a dipole and a quadrupole, similar to the one observed
in the VKS experiment. The agreement between the experimental results
and these numerical simulations show that, despite the high level of
turbulence and complexity of the flow, the generation of the magnetic
field can be understood using a few spatial properties of the
flow. Moreover, the mechanisms involved in the dynamics of the field
can be accurately described with a low dimensional model.

\acknowledgements I thank B. Gallet, N. Mordant, F. Petrelis, E. Dormy
and S. Fauve for useful discussions. Computations were performed at
CEMAG and CCRT centers.

\end{document}